\documentclass[preprint,12pt]{elsarticle} 
\usepackage{graphicx} 
\usepackage{amssymb} 

\journal{Nuclear Physics A} 

\begin{document} 

\begin{frontmatter} 

\title{Kaonic atoms and in-medium $K^- N$ amplitudes II: Interplay between 
theory and phenomenology} 

\author[a]{E.~Friedman} 
\author[a]{A.~Gal}  
\address[a]{Racah Institute of Physics, The Hebrew University, 
Jerusalem 91904, Israel} 

\date{\today} 

\begin{abstract} 

A microscopic kaonic-atom optical potential $V^{(1)}_{K^-}$ is constructed, 
using the Ikeda-Hyodo-Weise NLO chiral $K^-N$ subthreshold scattering 
amplitudes constrained by the kaonic hydrogen SIDDHARTA measurement, and 
incorporating Pauli correlations within the Waas-Rho-Weise generalization 
of the Ericson-Ericson multiple-scattering approach. Good fits to kaonic 
atom data over the entire periodic table require additionally sizable 
$K^-NN$--motivated absorptive and dispersive phenomenological terms, 
in agreement with our former analysis based on a post-SIDDHARTA in-medium 
chirally-inspired NLO separable model by Ciepl\'{y} and Smejkal. 
Such terms are included by introducing a phenomenological potential 
$V^{(2)}_{K^-}$ and coupling it self-consistently to $V^{(1)}_{K^-}$. 
Properties of resulting kaonic atom potentials are discussed with special 
attention paid to the role of $K^-$-nuclear absorption and to the extraction 
of density-dependent amplitudes representing $K^-$ multi-nucleon processes.  

\end{abstract}


\end{frontmatter}

\section{Introduction} 
\label{sec:intro}                                              

The wealth of strong-interaction data deduced from kaonic atoms provides 
invaluable information on the $K^-$-nuclear interaction at threshold 
\cite{BFG97,FG07}. Recent studies of $K^-$ atoms focused on constructing 
self-consistently a density-dependent $K^-$-nuclear potential: 
\begin{equation} 
2{\omega_K}V^{(1)}_{K^-}(\rho)=
-\:4\pi{\tilde{\cal F}}_{K^-N}(p_{\rho},\sqrt{s_{\rho}})\rho\;. 
\label{eq:t} 
\end{equation} 
Here $\omega_K \approx m_K$ at threshold, $\vec p_{\rho}$ is the in-medium 
relative $K^-N$ momentum and $s_{\rho}=(E_K+E_N)^2-({\vec p}_K+{\vec p}_N)^2$ 
is the in-medium Lorentz invariant Mandelstam energy-squared variable, both of 
which depend on the density $\rho$ \cite{CFG11a,CFG11b,FG12}. The in-medium 
scattering amplitude ${\tilde{\cal F}}_{K^-N}(p_{\rho},\sqrt{s_{\rho}})$ 
reduces in the low-density limit to 
${\tilde f}_{K^-N}$($p$=0,$\sqrt{s}$=$m_K$+$m_N$), 
the free-space $K^-N$ scattering length in the $K^-$ nucleus c.m. system. 
The dependence of ${\tilde{\cal F}}_{K^-N}(p_{\rho},\sqrt{s_{\rho}})$ 
on its kinematical variables was transformed in these studies, 
using a self-consistent procedure, into a density-dependent 
${\tilde{\cal F}}_{K^-N}(\rho)$. The resulting density-dependent 
optical potential $V^{(1)}_{K^-}(\rho)$ accounts for single-nucleon 
$\bar K N\to\bar K N$ elastic and $\bar K N\to\pi Y$ reaction processes. 
Empirically, $\bar K N$ amplitudes at about 40 MeV below threshold 
are involved in pinning down $V^{(1)}_{K^-}$ in kaonic atoms at half 
nuclear-matter density \cite{CFG11a,CFG11b,FG12}. 

This microscopic construction of $V^{(1)}_{K^-}$, however, did not provide 
any reasonable reproduction of the experimental values of strong-interaction 
level shifts and widths in kaonic atoms. In particular, because of the rapid 
decrease of the underlying absorptivity Im~${\tilde f}_{K^-N}(p,\sqrt{s})$ 
when the free-space amplitudes ${\tilde f}_{K^-N}$ are evaluated further below 
the $K^-N$ threshold, Im~$V^{(1)}_{K^-}$ was unable to account for the strong 
absorptivity content of kaonic atoms (i.e. their level widths). Thus, 
in addition to the single-nucleon terms on the r.h.s. of Eq.~(\ref{eq:t}), 
a sizable phenomenological absorptive term together with a strong dispersive 
term appeared necessary in order to achieve reasonable fits to the data. 
This has been demonstrated recently in Refs.~\cite{CFG11a,CFG11b,FG12} 
where in-medium $K^-N$ amplitudes ${\tilde{\cal F}}_{K^-N}$ constructed within 
the chirally motivated separable-interaction models of Ciepl\'{y} and Smejkal 
(CS) \cite{CS10,CS12} were used to evaluate $V^{(1)}_{K^-}$. In particular, 
models TW1 and NLO30 from Ref.~\cite{CS12}, accounting for the recently 
measured SIDDHARTA values of kaonic hydrogen $1s$ shift and width 
\cite{SID11,SID12}, have been used in Refs.~\cite{CFG11b} and \cite{FG12} 
respectively. It was found in these works that the added $V^{(2)}_{K^-}$ 
part consisting of dispersive and absorptive terms was as important as 
$V^{(1)}_{K^-}$. 

\begin{figure}[htb] 
\begin{center} 
\includegraphics[width=0.6\textwidth]{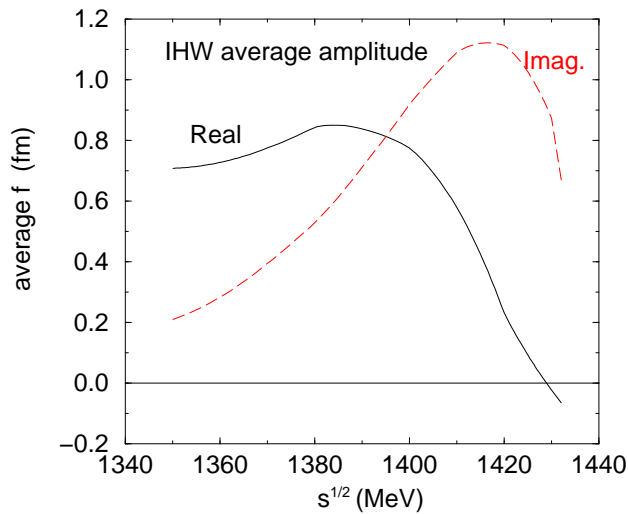} 
\caption{The $K^-N$ c.m. scattering amplitude $f_{K^-N}(\sqrt{s})=\frac{1}{2}
(f_{K^-p}(\sqrt{s})+f_{K^-n}(\sqrt{s}))$ below threshold, constructed from the 
IHW free-space $K^-p$ and $K^-n$ $s$-wave scattering amplitudes \cite{IHW12}. 
To convert from the two-body c.m. to the lab system, which for $A\gg 1$ 
coincides with the $K^-$ nucleus c.m. system, 
apply ${\tilde f}=(\sqrt{s}/m_N)f$.} 
\label{fig:IHWave} 
\end{center} 
\end{figure}

The present work is a natural extension of our recent work \cite{FG12} which 
may be considered part I of ongoing studies of kaonic atoms and in-medium 
$K^-N$ scattering amplitudes. The emphasis of the present work is on the 
interplay between theory and phenomenology that emerges in kaonic atom 
studies. Starting from the free-space NLO chiral $K^-p$ and $K^-n$ $s$-wave 
amplitudes constructed by Ikeda, Hyodo and Weise (IHW) \cite{IHW11,IHW12}
which account for the SIDDHARTA data, we arrive at similar conclusions to 
those reached in part I outlined above. We note that the IHW construction is 
free of any phenomenologically adjusted momentum-space form factors which in 
the CS separable-model construction are not directly guided by a systematic 
chiral hierarchy. The IHW free-space charge-averaged $K^-N$ c.m. scattering 
amplitude $f_{K^-N}(\sqrt{s})$ is shown in Fig.~\ref{fig:IHWave} below the 
$K^-N$ threshold. One notes its strong energy dependence, with 
Re~$f_{K^-N}(\sqrt{s})$ mostly rising in going subthreshold and 
Im~$f_{K^-N}(\sqrt{s})$ sharply dropping below $\sqrt{s}=1415$~MeV as one 
approaches the $\pi\Sigma$ threshold ($\sqrt{s}=1330$~MeV). 

To generate in-medium amplitudes ${\tilde{\cal F}}_{K^-N}(\sqrt{s_{\rho}})$ 
from the IHW free-space $K^-p$ and $K^-n$ $s$-wave amplitudes, we apply the 
multiple-scattering (MS) approach of Waas, Rho and Weise (WRW) \cite{WRW97} 
focusing on Pauli correlation effects, as described in Section~\ref{sec:meth}. 
Charge (isospin) degrees of freedom are incorporated in this MS approach which 
determines, under a straightforward generalization of Eq.~(\ref{eq:t}), the 
single-nucleon density-dependent potential $V^{(1)}_{K^-}(\rho)$. To represent 
multi-nucleon dispersive and absorptive processes, we add a phenomenological 
density-dependent interaction potential $V^{(2)}_{K^-}(\rho)$. 
Both $V^{(1)}_{K^-}$ and $V^{(2)}_{K^-}$ are coupled implicitly within 
a self-consistent cycle built into the kaonic atom fitting procedure. 
Properties of the resulting kaonic atom potentials are discussed in 
Section~\ref{sec:Res} with special emphasis placed on the role of $K^-$ 
nuclear absorption. 
The paper ends with a brief summary in Section~\ref{sec:sum}.

\section{Methodology} 
\label{sec:meth} 

Several subjects are introduced and discussed in this section. 
In Section~\ref{subsec:wrw} we briefly review the multiple-scattering 
procedure applied by WRW \cite{WRW97} to incorporate nuclear-correlation 
contributions, particularly Pauli correlations in the construction of a $K^-$ 
nucleus potential $V^{(1)}_{K^-}$ at low energies in terms of in-medium 
$K^- N$ amplitudes ${\tilde{\cal F}}_{K^-N}(\sqrt{s_{\rho}})$. In Section 
\ref{subsec:V2} we introduce a phenomenological potential $V^{(2)}_{K^-}$ to 
represent multi-nucleon processes outside of the underlying meson-baryon chiral 
model on which $V^{(1)}_{K^-}$ is based. Although no explicit coupling between 
$V^{(1)}_{K^-}$ and $V^{(2)}_{K^-}$ is practised in the present calculations, 
in agreement with the spirit of the original Ericson-Ericson MS procedure 
\cite{EE66}, possible alternatives are considered in Appendix A to this work.  
In Section~\ref{subsec:Evsrho} we focus attention on the self-consistent 
procedure of relating subthreshold energies to densities through functional 
dependence involving {\it both} $V^{(1)}_{K^-}$ and $V^{(2)}_{K^-}$, 
thereby coupling implicitly $V^{(1)}_{K^-}$ to $V^{(2)}_{K^-}$. 
Finally, in Section~\ref{subsec:L1405} we discuss the effect of the 
$\Lambda(1405)$ subthreshold resonance on the low-density behavior of 
our in-medium amplitudes ${\tilde{\cal F}}_{K^-N}(\rho)$.

\subsection{Overview of WRW} 
\label{subsec:wrw}

In this MS procedure one starts by relating the meson wavefunction 
$\phi(\vec r)$ generated by the incident plane wave 
$\exp(i{\vec k}\cdot{\vec r})$ to the effective meson wavefunctions 
$\phi_{I}^{\rm eff}(\vec {r'})$ at the scattering point $\vec {r'}$: 
\begin{equation} 
\phi(\vec r)=\exp({\rm i}{\vec k}\cdot{\vec r})+\sum_{I=0,1}\int{d^{3}r'\;
\frac{\exp({\rm i}k|{\vec r}-{\vec {r'}}|)}{|{\vec r}-{\vec {r'}}|}\;
{\tilde f}_{I}\rho_{I}(\vec {r'})\phi_{I}^{\rm eff}(\vec {r'})}\;. 
\label{eq:wrw6} 
\end{equation} 
Here, ${\tilde f}_{I}$ are free-space $\bar K N$ $s$-wave scattering 
amplitudes with good isospin in the lab system and $\rho_{I}=(2I+1)\rho/4$ 
for isospin-symmetric nuclear density $\rho$. Operating with $(\Delta +k^2)$ 
on both sides of (\ref{eq:wrw6}) one obtains the wave equation 
\begin{equation} 
(\Delta + k^2)\;\phi(\vec r)=-4\pi\sum_{I=0,1}{\tilde f}_{I}
\rho_{I}(\vec r)\phi_{I}^{\rm eff}(\vec r) \;. 
\label{eq:wrw7} 
\end{equation} 
The MS procedure relates effective wavefunctions $\phi_{I}^{\rm eff}(\vec r)$ 
at $\vec r$ to effective wavefunctions $\phi_{I'}^{\rm eff}(\vec {r'})$ at 
$\vec {r'}$ as follows: 
\begin{equation} 
\phi_{I}^{\rm eff}(\vec r)=\phi(\vec r)+\sum_{I'=0,1}\int{d^{3}r'\;\frac
{\exp({\rm i}k|{\vec r}-{\vec {r'}}|)}{|{\vec r}-{\vec {r'}}|}\;c_{II'}
({\vec r},{\vec {r'}}){\tilde f}_{I'}\rho_{I'}(\vec {r'})
\phi_{I'}^{\rm eff}(\vec {r'})} \;, 
\label{eq:wrw9} 
\end{equation} 
where $c_{II'}({\vec r},{\vec {r'}})$ are isospin projected $NN$ correlation 
functions which in nuclear matter depend on $t=|{\vec r}-{\vec {r'}}|$ only. 
Replacing in the long-wavelength limit the argument $\vec {r'}$ by $\vec r$ in 
$\rho_{I'}(\vec {r'})$ and $\phi_{I'}^{\rm eff}(\vec {r'})$, 
Eq.~(\ref{eq:wrw9}) reduces to 
\begin{equation} 
\phi_{I}^{\rm eff}(\vec r)=\phi(\vec r)-\sum_{I'=0,1}\xi_{II'}
{\tilde f}_{I'}\rho_{I'}(\vec r)\phi_{I'}^{\rm eff}(\vec r) \;, 
\label{eq:wrw11} 
\end{equation} 
where 
\begin{equation} 
\xi_{II'}=-4\pi\int_0^{\infty}{dt\exp(ikt)\,t\,c_{II'}(t)} \;. 
\label{eq:wrw12} 
\end{equation} 
The following discussion is limited to Pauli correlation contributions to 
$c_{II'}$ which were found by WRW (and also confirmed by us) to overshadow 
contributions from dynamical short-range correlations. For isospin symmetric 
matter, the Pauli $\xi_{II'}$ are diagonal in isospin, see Ref.~\cite{WRW97}. 
Solving Eqs.~(\ref{eq:wrw11}) for $\phi_{I}^{\rm eff}(\vec r)$ in terms 
of $\phi(\vec r)$ and substituting on the r.h.s. of Eq.~(\ref{eq:wrw7}), 
the latter assumes the form of 
$2{\omega_K}V^{(1)}_{K^-}(\omega_K;\rho)\;\phi(\vec r)$, with  
\begin{equation} 
2{\omega_K}V^{(1)}_{K^-}(\omega_K;\rho)=-\:4\pi\:\sum_{I}\frac{2I+1}{4}\;
\frac{{\tilde f}_{I}}{1+\frac{1}{4}\xi_k{\tilde f}_{I}\rho(r)}\:\rho(r) \;, 
\label{eq:wrw26} 
\end{equation} 
where $\xi_k$ is given by 
\begin{equation} 
\xi_k=\frac{9\pi}{p_F^2}\left( 4\int_0^{\infty}{\frac{dt}{t}\,\exp({\rm i}qt)\,
j_1^2(t)} \right) \; , \;\;\;\;\;\;\; q=k/p_F \;. 
\label{eq:wrw25} 
\end{equation} 
For kaonic atoms, $k=(\omega_K^2-m_K^2)^{1/2}\approx 0$ 
and $\xi_{k=0}=9\pi/p_F^2$, where the Fermi momentum $p_F$ is given by 
$p_F=(3{\pi}^2\rho/2)^{1/3}$. We note that the density dependence of 
$V^{(1)}_{K^-}(\omega_K;\rho)$ in Eq.~(\ref{eq:wrw26}) is not limited to the 
explicit dependence on $\rho$ in its right-hand side, but arises as well from 
the $\sqrt{s}$ energy argument of the free-space amplitudes ${\tilde f}_{I}$ 
which for simplicity was disregarded in the derivation above and which 
gives rise in the nuclear medium to a density-dependent subthreshold energy 
argument $\sqrt{s_{\rho}}$. The precise definition of $\sqrt{s_{\rho}}$ 
and the self-consistent procedure by which the $\sqrt{s_{\rho}}$ dependence 
of ${\tilde f}_{I}$ is converted into a density dependence of 
${\tilde{\cal F}}_{K^-N}$ and of $V^{(1)}_{K^-}$ are relegated 
to Section~\ref{subsec:Evsrho}. 

Eq.~(\ref{eq:wrw26}) was derived assuming implicitly equal proton and 
neutron density distributions, $\rho_p(r)$=$\rho_n(r)$=$\rho(r)/2$. Relaxing 
this constraint, primarily for use in kaonic atom applications where 
$\omega_K$$\approx$$\mu_K$ with $\mu_K$ the reduced mass of the kaon, one 
obtains to leading order: 
\begin{equation} 
2{\mu_K}V^{(1)}_{K^-}(\rho)=-\:4\pi\:\left[ \frac{(2{\tilde f}_{K^-p}-
{\tilde f}_{K^-n})\:\frac{1}{2}\rho_p}{1+\frac{1}{4}\xi_{k=0}{\tilde f}_0
\rho(r)}+\frac{{\tilde f}_{K^-n}(\frac{1}{2}\rho_p+\rho_n)}
{1+\frac{1}{4}\xi_{k=0}{\tilde f}_1\rho(r)}\right]\;. 
\label{eq:wrw14} 
\end{equation} 
This form of the MS summation is used in all of the numerical applications 
reported in the present work. Eq.~(\ref{eq:wrw14}) may be rewritten as 
\begin{equation} 
2{\mu_K}V^{(1)}_{K^-}(\rho)=-\:4\pi\:(1+\frac{A-1}{A}\frac{\mu_K}{m_N})\:
[{\cal F}_{K^-p}(\rho)\rho_p(r) + {\cal F}_{K^-n}(\rho)\rho_n(r)]\;,
\label{eq:calFpn} 
\end{equation} 
which defines the in-medium amplitudes ${\cal F}_{K^-p}(\rho)$ and 
${\cal F}_{K^-n}(\rho)$. Here, $A$ is the atomic mass number. Finally, to 
bring Eq.~(\ref{eq:calFpn}) into the more compact form (\ref{eq:t}), we 
define an {\it effective} amplitude ${\cal F}_{K^-N}^{\rm eff}(\rho)$ through 
\begin{equation} 
{\cal F}_{K^-N}^{\rm eff}(\rho)\rho(r) =
{\cal F}_{K^-p}(\rho)\rho_p(r) + {\cal F}_{K^-n}(\rho)\rho_n(r)\;,  
\label{eq:effA} 
\end{equation} 
with the standard conversion from two-body c.m. to the $K^-$-nuclear 
c.m. frame given by ${\tilde{\cal F}}_{K^-N}^{\rm eff}(\rho)=
(1+\frac{A-1}{A}\frac{\mu_K}{m_N}){\cal F}_{K^-N}^{\rm eff}(\rho)$. 
Eq.~(\ref{eq:effA}) may serve for defining similarly an effective 
free-space subthreshold amplitude $f_{K^-N}^{\rm eff}(\rho)$, with 
density dependence arising from the underlying energy dependence 
$\sqrt{s}\to\sqrt{s_{\rho}}\to\rho$.

\subsection{Adding $V^{(2)}_{K^-}(\omega_K;\rho)$} 
\label{subsec:V2} 

We wish now to account for multi-nucleon dispersive and absorptive processes 
by adding a phenomenological term $V^{(2)}_{K^-}(\omega_K;\rho)$ to the MS 
single-nucleon potential $V^{(1)}_{K^-}(\omega_K;\rho)$, Eq.~(\ref{eq:wrw26}). 
Traditionally, these processes are not iterated within the MS expansion in 
which scattering occurs on single nucleons via amplitudes $\tilde f$, and 
$V^{(2)}_{K^-}$ is simply {\it added} to the MS $V^{(1)}_{K^-}$ specified 
in Eq.~(\ref{eq:wrw26}) by $\xi_{k=0}=9\pi/p_F^2$ for kaonic atoms. 
Alternative ways of introducing $V^{(2)}_{K^-}$, by letting it affect 
explicitly the MS procedure for deriving the in-medium $V^{(1)}_{K^-}$, 
or by redefining the splitting of $V_{K^-}$ into its $V^{(1)}_{K^-}$ and 
$V^{(2)}_{K^-}$ components, are discussed briefly in Appendix A to this work. 
None of these alternatives was found in the present work to offer advantage 
over the straightforward addition of $V^{(2)}_{K^-}$. 

\subsection{Energy vs. density} 
\label{subsec:Evsrho} 

Here we outline a basic difference between the present work and our previous 
ones \cite{CFG11a,CFG11b,FG12} in which a model of using $K^-N$ amplitudes 
below threshold was introduced. In these calculations, energies and momenta 
of the $K^-$ meson and a bound nucleon were determined, independently, 
by the nuclear environment, and the Mandelstam energy variable $\sqrt{s}$ 
was evaluated in the nuclear medium, thereby becoming density dependent, 
$\sqrt{s_{\rho}}$. In particular, the in-medium $K^-$ momentum was assigned 
locally to the real part of the single-nucleon $K^-$-nucleus potential 
Re~$V^{(1)}_{K^-}$, leading to 
\begin{equation} 
\sqrt{s_{\rho}} \approx E_{\rm{th}} - B_N - \xi_NB_K -15.1(\rho/\rho_0)^{2/3} 
 +\xi_K({\rm Re}~V^{(1)}_{K^-}+V_c)~~({\rm MeV}) 
\label{eq:SC} 
\end{equation} 
with $E_{\rm{th}}=m_N+m_K,~\xi_N=m_N/(m_N+m_K),~\xi_K = m_K/(m_N+m_K)$, 
and with $V_c$ for the $K^-$ Coulomb potential, $B_N$ for a nucleon average 
binding energy and $\rho_0$ for nuclear matter density. The atomic binding 
energy $B_K$ of the $K^-$ is relatively small and, hence, it is safe to 
neglect it in Eq.~(\ref{eq:SC}). With a fixed value for $B_N$ it is evident 
that for $\rho \rightarrow 0$ the energy approaches $E_{\rm{th}}-B_N$, thus 
violating the low-density limit whereby the 1$N$ amplitude should approach 
the free amplitude at threshold, $E_{\rm{th}}$. In the present work we 
have therefore replaced the fixed average nucleon binding energy $B_N$ 
by a density-dependent one, 
\begin{equation} 
B_N(\rho)=B_N \rho /{\bar \rho} \; , 
\label{eq:Brho} 
\end{equation} 
where the average density is given by 
\begin{equation} 
{\bar \rho}=\frac{1}{A} \int{\rho ^2\;d^3r} 
\label{eq:rhobar} 
\end{equation} 
with $A$ the atomic mass number. For $B_N$ we used the value of 8.5~MeV, 
the same as in our earlier work. Furthermore, in order to enhance the very 
slow convergence of $\sqrt{s_{\rho}}$ to $E_{\rm th}$ due to the Coulomb 
potential, $V_c$ was multiplied in Eq.~(\ref{eq:SC}) by $(\rho/\rho_0)^{1/3}$, 
based on dimensional arguments, thereby ensuring that the low-density limit is 
respected in the calculations. 

Since $V^{(1)}_{K^-}$ is proportional locally according to Eq.~(\ref{eq:t}) 
to ${\tilde{\cal F}}_{K^-N}(\sqrt{s_{\rho}})$, and $\sqrt{s_{\rho}}$ according 
to Eq.~(\ref{eq:SC}) depends on Re~$V^{(1)}_{K^-}$, a self-consistent (SC) 
procedure was applied, with 5--6 iterations proving more than adequate for 
convergence. In this way a simple algorithm for constructing $V^{(1)}_{K^-}$ 
from any given model for ${\tilde{\cal F}}_{K^-N}(\sqrt{s_{\rho}})$ was 
formulated. As mentioned in the Introduction, the SC $V^{(1)}_{K^-}$ 
potentials constructed thereby were characterized by ignoring multi-nucleon 
absorption processes and by an imaginary part that goes rapidly to zero 
towards the $\pi\Sigma$ threshold. There were no free parameters at this 
phase of the calculation and there was no coupling between the real and 
imaginary parts of the potential beyond that provided by the input $K^-N$ 
amplitudes. In carrying out global fits to strong-interaction shifts and 
widths data across the periodic table, a phenomenological potential 
$V^{(2)}_{K^-}$ was added to the SC $V^{(1)}_{K^-}$, with parameters 
determined in a $\chi ^2$ fit search without perturbing the prior SC 
determination of $V^{(1)}_{K^-}$. However, no compelling reason was given 
why $V^{(2)}_{K^-}$ was excluded from the SC procedure. 

In the present work, a phenomenological term $V^{(2)}_{K^-}$ is included 
from the outset in the SC procedure. Eq.~(\ref{eq:SC}) is thus modified, 
replacing $V^{(1)}_{K^-}$ by $V_{K^-}=V^{(1)}_{K^-}+V^{(2)}_{K^-}$: 
\begin{equation} 
(\sqrt{s_{\rho}})_{\rm atom}\approx E_{\rm{th}}-B_N\rho/{\bar \rho}
-15.1(\rho/\rho_0)^{2/3} 
 +\xi_K({\rm Re}~V_{K^-}+V_c (\rho/\rho_0)^{1/3}) \;\;\; ({\rm MeV}) \; , 
\label{eq:SCmodif} 
\end{equation} 
where the subscript {\it atom} indicates the limitation to kaonic atoms, 
thereby also setting $B_K$=0. The first effect of this modification is to 
introduce {\it implicit coupling} between $V^{(1)}_{K^-}$ and $V^{(2)}_{K^-}$, 
since varying parameters of $V^{(2)}_{K^-}$ affects the $\sqrt{s_{\rho}}$ 
energy argument of the in-medium ${\tilde{\cal F}}_{K^-N}(\sqrt{s_{\rho}})$ 
and that of the underlying free-space ${\tilde f}_{K^-N}(\sqrt{s_{\rho}})$ 
amplitudes. Early tests of this approach using the NLO30 amplitudes revealed 
\cite{FG12} that this coupling is non-negligible when the imaginary part of 
the single-nucleon amplitude drops sharply below threshold, which was 
particularly the case with the `SE' in-medium version of model NLO30. 
The increased flexibility due to this coupling leads in the present 
work, building on the IHW free-space amplitudes, to good fits to the data, 
well beyond what was achieved in our earlier works \cite{CFG11a,CFG11b,FG12}. 

\begin{figure}[htb] 
\begin{center} 
\includegraphics[width=0.6\textwidth]{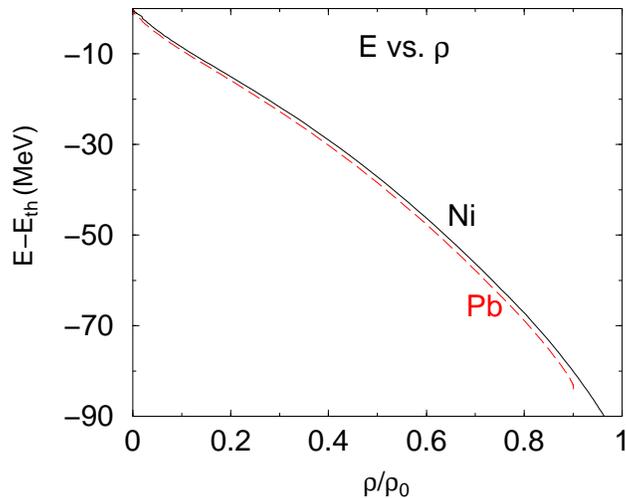} 
\caption{Subthreshold $K^-N$ energies probed by the $K^-$ nuclear potential 
at threshold as a function of nuclear density in Ni and Pb, calculated 
self-consistently according to Eq.~(\ref{eq:SCmodif}) from the IHW-based 
global fit to kaonic atoms specified in Section~\ref{sec:Res}, see text.} 
\label{fig:WRW12Evsrho} 
\end{center} 
\end{figure} 

Fig.~\ref{fig:WRW12Evsrho} shows the functional dependence generated 
by the SC relation Eq.~(\ref{eq:SCmodif}) between subthreshold $K^-N$ 
energies and nuclear densities for Ni and Pb, calculated using the 
IHW-based global fit to kaonic atoms detailed in Section~\ref{sec:Res}. 
Compared with the earlier version \cite{FG12} where only $V^{(1)}_{K^-}$ 
appeared in the SC expression, Eq.~(\ref{eq:SC}), lower energies are now 
probed at the higher densities and higher energies are probed at lower 
densities. Consequently, for the more relevant region of 50\% of the central 
nuclear density, the energy downward shift remains unchanged at 40~MeV below 
threshold.

\subsection{Features of density dependence} 
\label{subsec:L1405} 

\begin{figure}[htb]
\begin{center}
\includegraphics[width=0.6\textwidth]{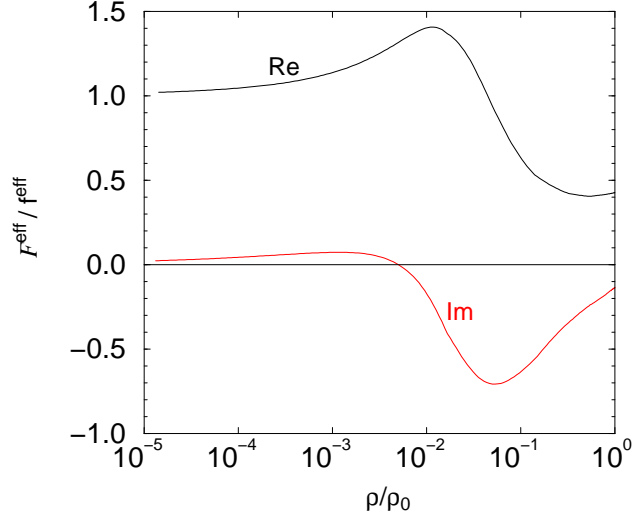}
\caption{Ratio of the in-medium effective IHW amplitude Eq.~(\ref{eq:effA})
to the free effective IHW amplitude as a function of nuclear density in Ni.}
\label{fig:compxratio}
\end{center}
\end{figure}

Having introduced by Eq.~(\ref{eq:SCmodif}) a relationship between the 
subthreshold energy argument of $K^-N$ amplitudes and their implied 
density dependence, it is instructive to demonstrate the density 
dependence generated by the WRW renormalization Eq.~(\ref{eq:wrw14}). 
Shown in Fig.~\ref{fig:compxratio} is the ratio $\cal R$ of the in-medium 
effective amplitude ${\cal F}_{K^-N}^{\rm eff}$, Eq.~(\ref{eq:effA}), to 
the free effective amplitude $f_{K^-N}^{\rm eff}$, calculated as a function 
of nuclear density in Ni for the IHW amplitudes in the absence of the 
$V^{(2)}_{K^-}$ potential term. Note the logarithmic density scale used 
to highlight the slow convergence of $\cal R$ to its low-density limit. 
For example, Re~$\cal R$=1.14 and Im~$\cal R$=0.07 near 0.1\% of the Ni 
central density $\rho_0$, still far from their limiting values Re~$\cal R$=1 
and Im~$\cal R$=0, respectively. This slow convergence is caused by the 
predominance of the $\Lambda(1405)$ resonance for densities roughly 
below 6\% of $\rho_0$ where Re~$\cal R$ exhibits hump structure with 
values exceeding 1, owing to the large negative values assumed by 
Re~${\tilde f}_{I=0}$ near threshold; the position of the $\Lambda(1405)$ 
is marked here by the vanishing of Re~${\tilde f}_{I=0}$ at 
$\sqrt{s}$$\approx$1415~MeV. At densities above 6\% of $\rho_0$, 
Re~${\cal R}{\leq}1$, decreasing monotonically with density owing to the 
rapid increase of Re~${\tilde f}_{I=0}$ below 1415~MeV and leveling off 
when Re~${\tilde f}_{I=0}$ has reached its (positive) maximum value. 

A similar analysis shows that the $\Lambda(1405)$ is directly responsible 
for Im~$\cal R$ becoming slightly positive below a fraction of 1\% of 
$\rho_0$ and slowly converging to its low-density limiting value of 0. 
However, it is not clear whether the isospin-based WRW method of introducing 
medium effects into the free-space IHW amplitudes, as applied here, is valid 
at such low densities where the $K^-$ Coulomb potential $V_c$ becomes 
comparable with and even exceeds $V^{(1)}_{K^-}$. 
In order to check sensitivities, $\chi ^2$ fits were made also by imposing 
a transition to the free-space amplitudes at densities of 0.5, 1.0 
and 2.0\% of $\rho_0$, while retaining $V_{K^-}$ down to densities 
$\approx$10$^{-5}\rho_0$ as normally done in kaonic atom calculations. 
Below 10$^{-5}\rho_0$, only $V_c$ is retained, along with changing the radial 
integration scheme from nuclear to atomic dimensions. Weak sensitivities 
were found to the way the amplitudes were handled at very low densities. 
Most of the results in this paper are for transition to free amplitudes 
at 2\% relative density.

\section{Results and discussion} 
\label{sec:Res} 

Here we report and discuss results of fits to kaonic atom data, using the 
methodology outlined in the previous section. A phenomenological potential 
$V^{(2)}_{K^-}$ representing $K^-$ multi-nucleon processes is defined in 
Section~\ref{subsec:WRW12}, the combined amplitudes that result from the 
self-consistent fitting procedure are shown and discussed, and the emergence 
of deep $K^-$ nucleus potentials is noted. The present IHW-based $K^-$ nuclear 
potential is compared in Section~\ref{subsec:comparisons} with the NLO30 
potentials of Ref.~\cite{FG12} and, in view of the significance of the 
resulting $V^{(2)}_{K^-}$ component, its signatures are discussed in 
Section~\ref{subsec:separation} with special emphasis on its absorptive 
part.

\subsection{IHW-based potentials} 
\label{subsec:WRW12} 

Fits to 65 data points for strong-interaction observables in kaonic atoms 
were made by introducing a phenomenological multi-nucleon term 
\begin{equation} 
V^{(2)}_{K^-}=-(4\pi/2\mu_K )[b\rho+B\rho(\rho/\rho_0)^\alpha] \; , 
\label{eq:phen} 
\end{equation} 
where $\mu_K$ is the kaon-nucleus reduced mass, $b$ and $B$ are 
energy-independent complex parameters and the real exponent $\alpha$ satisfies 
$\alpha > 0$. The linear density term with coefficient $b$ could be viewed as 
compensating for uncertainties in the IHW free-space amplitude input. 
The combined potential $V^{(1)}_{K^-}+V^{(2)}_{K^-}$ was calculated according 
to the prescriptions of Section~\ref{sec:meth} and fits to the data were made 
varying the parameters of $V^{(2)}_{K^-}$. The real part of the {\it full} 
potential $V^{(1)}_{K^-}+V^{(2)}_{K^-}$ was used in the present SC procedure 
Eq.~(\ref{eq:SCmodif}). Since good convergence was limited to varying up 
to four parameters, we varied the parameters $b$ and $B$ while griding on the 
parameter $\alpha$. Strong correlations were observed between the parameters 
Im~$b$ and $\alpha$ and the lowest $\chi^2$ of 107 for 65 data points was 
obtained for Im~$b$=$-$1.11$\pm$0.12~fm and $\alpha$=0.3. However, this 
magnitude of $\mid$Im~$b$$\mid$ is too large to absorb into the uncertainties 
allowed by the IHW determination of the free-space $\bar K N$ scattering 
amplitudes at threshold \cite{IHW12}. Constraining Im~$V^{(2)}_{K^-}$ to be 
non-negative at all values of $\rho$, we set Im~$b$=0 resulting in 
$\chi^2$=118 for $\alpha$=1.2, Re~$b$=$-$0.34$\pm$0.07~fm, 
Re~$B$=1.94$\pm$0.16~fm and Im~$B$=0.83$\pm$0.16~fm. This solution is used in 
the rest of this work. We note that the value of Re~$b$ could be related to 
a decrease of the poorly known Re~$a_{K^-n}$ by about 0.45~fm, twice as much 
as the downward uncertainty assigned in the IHW NLO determination. 

\begin{table} 
\caption{Characteristic quantities of $K^-$Ni potentials from global fits 
to kaonic atom data. DD is a purely phenomenological (density-dependent) 
model, NLO30 is the model used in our previous work \cite{FG12} and IHW 
is the present model based on the free-space input from Ref.~\cite{IHW12}. 
Values of potentials are in MeV, values of r.m.s. radii are in fm. 
The r.m.s. radius of the point-proton distribution in Ni is 3.69 fm.} 
\label{tab:res} 
\begin{center} 
\begin{tabular}{ccccccc} 
\hline 
model&$\chi ^2$ (N=65)&$V_{\rm R}(0)$&$V_{\rm I}$(0)& 
$r_{\rm R}$&$r_{\rm I}$ & $\alpha$ \\ \hline 
DD&103&$-$199&$-$76&3.48&3.71& 0.25\\ 
IHW&118&$-$191&$-$79&3.34&3.73 & 1.2\\ 
NLO30&148&$-$179&$-$71&3.42&3.70 & 1 \\ \hline 
\end{tabular} 
\end{center} 
\end{table} 

Table~\ref{tab:res} summarizes the results of fits to the data.
The DD potential listed in the table is a purely 
phenomenological potential of a 
form similar to Eq.~(\ref{eq:phen}) and is included as a `benchmark' for what 
could be an ultimate fit to 65 data points across the periodic table, where 
the data come from several experiments at different laboratories. 
The value of $\chi ^2$=103 for 65 data points ($\chi ^2$/dof of 1.7) is most 
acceptable. The entry for the NLO30 model \cite{FG12} is typical of results 
obtained in the recent works \cite{CFG11a,CFG11b,FG12}. As was shown already 
in Ref.~\cite{FGB94} and re-iterated recently, the r.m.s. radii of the 
potentials are also characteristic quantities that 
reflect the density dependence 
of the potentials.
                    
\begin{figure}[thb]
\begin{center}
\includegraphics[width=0.6\textwidth]{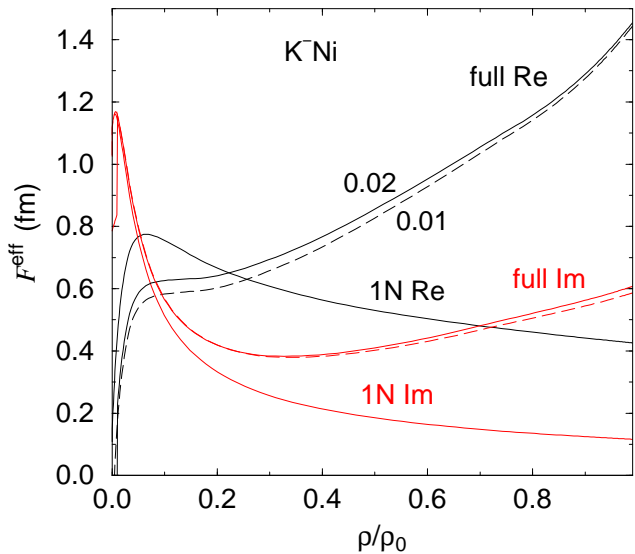}
\caption{Effective $K^-N$ full amplitudes as a function of nuclear density in 
Ni, based on the in-medium IHW 1$N$ amplitudes (also shown). Solid curves are 
for matching to the free-space amplitudes at 0.02 of the nuclear density, 
dashed curves are for matching at 0.01 relative density, see text.}  
\label{fig:fullampl}
\end{center}
\end{figure}

Fig.~\ref{fig:fullampl} shows the full effective $K^-N$ amplitudes 
${\cal F}^{\rm eff}(\rho)$ and the corresponding in-medium IHW-based 
single-nucleon (1$N$) amplitudes for a typical nucleus, $^{58}$Ni. 
The increase or decrease of the amplitude with density leads to decrease 
or increase, respectively, of the radii of the corresponding part of the 
potential relative to the nuclear radius, in agreement with the r.m.s. radii 
listed in Table~\ref{tab:res}. The monotonic increase with density of the 
real part of the full effective amplitude ${\cal F}^{\rm eff}(\rho)$ 
is reflected in the relatively small r.m.s. radius of the real potential. 
This increase contrasts with the decrease with density of the in-medium 
1$N$ real amplitude above 10$\%$ of central nuclear density. 
On the other hand, for the imaginary part of the full effective amplitude 
${\cal F}^{\rm eff}(\rho)$, the combined effect of decrease followed 
by increase with density causes the r.m.s. radius of the imaginary potential 
to be very close to 3.69~fm which is the radius of the proton distribution. 
In contrast, the imaginary part of the in-medium 1$N$ amplitude practically 
decreases with density. The figure shows how the difference between the 
imaginary parts of the full amplitude and its 1$N$ component increases 
steadily with density, providing solid evidence for the dominance of 
multi-nucleon $K^-$-nuclear absorption at $\rho/\rho_0 \geq 0.5$. 

\begin{figure}[thb] 
\begin{center} 
\includegraphics[width=0.6\textwidth]{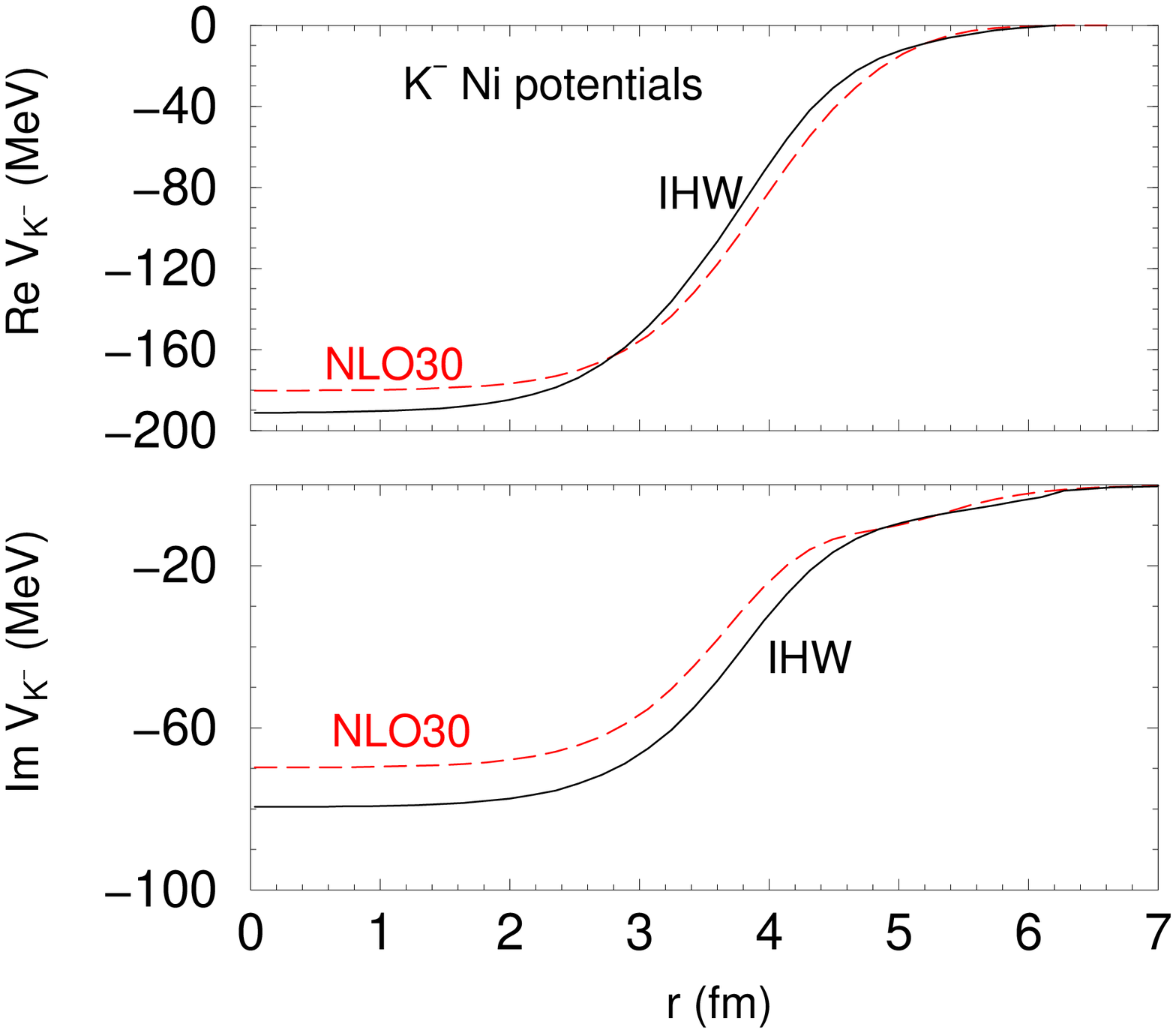} 
\caption{$K^-$ nuclear potentials for $K^-$ atoms of Ni.
Solid curves: derived from global fits based on 
in-medium IHW amplitudes (plus a phenomenological term). 
Dashed curves: derived from in-medium NLO30 amplitudes 
(plus a phenomenological term) \cite{FG12}.} 
\label{fig:WRW1Nipotl} 
\end{center} 
\end{figure} 

\begin{figure}[thb] 
\begin{center} 
\includegraphics[width=0.6\textwidth]{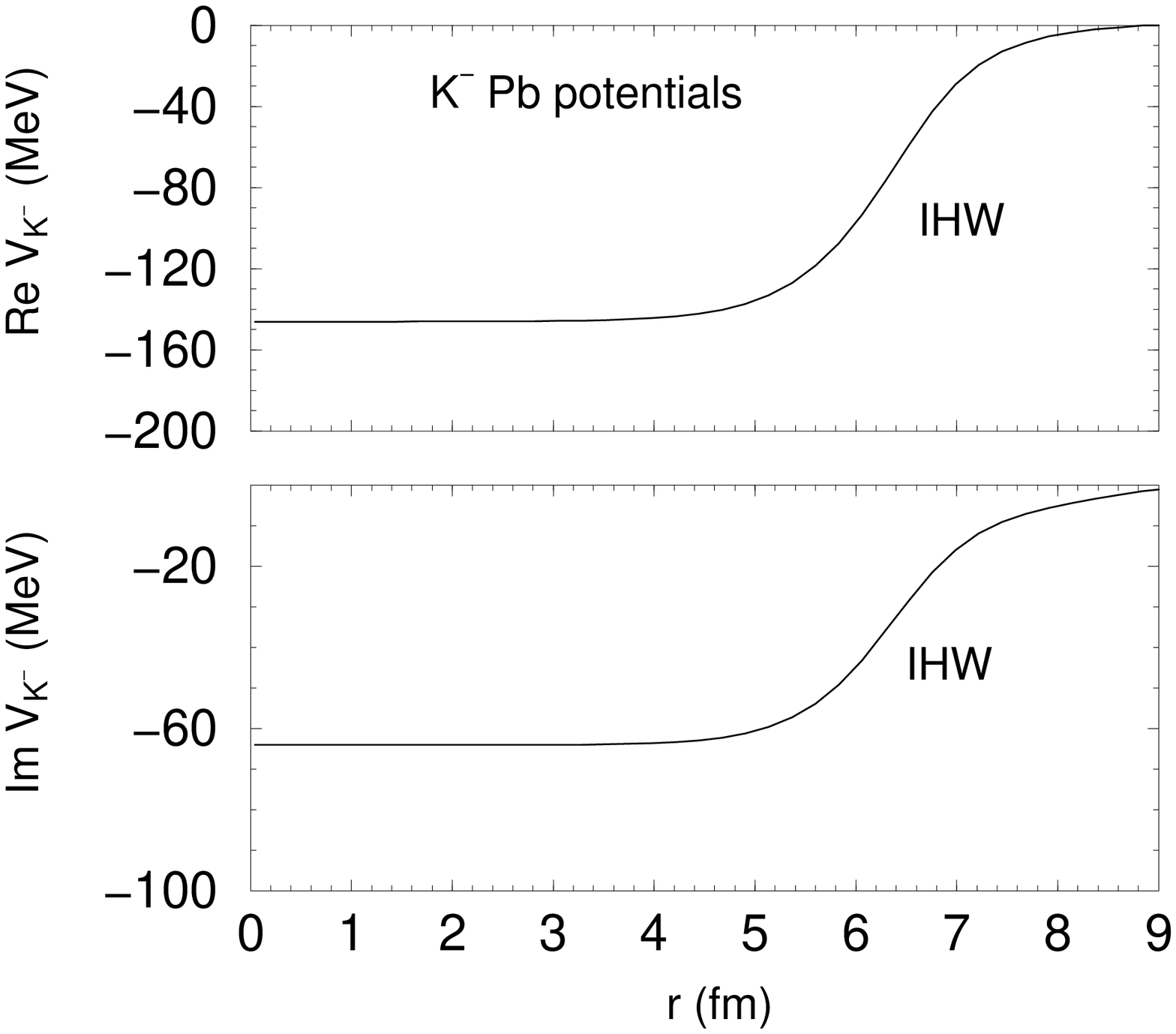} 
\caption{$K^-$ nuclear potentials for $K^-$ atoms of Pb, 
derived from global fits based on in-medium IHW amplitudes 
(plus a phenomenological term).} 
\label{fig:WRW1Pbpotl} 
\end{center} 
\end{figure} 

Finally, Figs.~\ref{fig:WRW1Nipotl} and \ref{fig:WRW1Pbpotl} show the full 
IHW-based potential for Ni and Pb, respectively, where the NLO30 in-medium 
potential (plus a phenomenological term) \cite{FG12} is also shown for Ni. 
Note the sizable attraction of order 150--200~MeV at central densities. 
These potential depths obviously reflect a smooth extrapolation provided 
by the density dependence of the IHW-based $V^{(1)}_{K^-}$ and that of 
$V^{(2)}_{K^-}$, Eq.~(\ref{eq:phen}). However, depths of order 80--90~MeV 
around the half-density radius and down to 1~fm inward are reliably determined 
in kaonic atom fits \cite{BFr07}, and the depths deduced in the present work 
are fully consistent with those deduced in earlier studies, beginning with 
the DD phenomenological analysis from 1994 \cite{FGB94} and ending with the 
very recent NLO30 theory-based analysis \cite{FG12}.

\subsection{Comparison with previous results}
\label{subsec:comparisons}

Fig.~\ref{fig:WRW1Nipotl} shows a comparison of the present IHW-based 
$K^-$Ni potential with the corresponding potential for the NLO30 model. 
The real potentials are quite similar in the two cases whereas the imaginary 
potentials are different. It is instructive to trace the origin of the very 
significant difference of 30 units in $\chi ^2$ values between the two models 
as shown in Table \ref{tab:res}. 
Inspecting the fits for the different target nuclei reveals that most of the 
reduction in $\chi ^2$ in the present work comes  
from species where strong-interaction observables were measured for more 
than a single kaonic atom level. 

Shifts and widths are usually measured for a `lower' level and in favourable 
cases a relative yield of the transition from the next higher level to the 
lower level can also be measured. That quantity provides the width of the 
`upper' level, based on the known radiation width for the electromagnetic 
transition. The smaller values of $\chi ^2$ achieved in the present work are 
due to better agreement, in most cases, between calculation and experiment 
for widths of lower and upper levels in the {\it same} kaonic atom.
Note that strong-interaction effects in kaonic atoms are dominated by
absorption processes \cite{BFG97,FG07}. Indeed, strong-interaction level 
widths in kaonic atoms are up to one order of magnitude larger than the 
corresponding strong-interaction level shifts. Furthermore, these shifts 
are almost universally repulsive, although the real potentials required 
to fit kaonic atom data are attractive. It is therefore not surprising 
that the differences between the imaginary potentials are the origin of 
the significantly better fits achieved in the present work. 

\begin{figure}[thb] 
\begin{center} 
\includegraphics[width=0.6\textwidth]{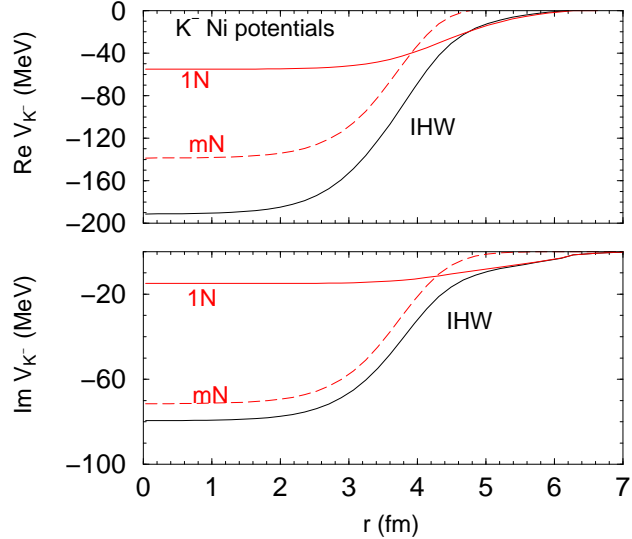} 
\caption{Components of the $K^-$ nuclear potentials for $K^-$ atoms of Ni, 
with the 1$N$ component derived from IHW amplitudes, see text.} 
\label{fig:WRW1nonadd} 
\end{center} 
\end{figure}

\subsection{On separating $V^{(2)}_{K^-}$ from $V^{(1)}_{K^-}$} 
\label{subsec:separation} 

In view of the dominance of absorption processes, it is instructive to study 
the relative importance of multi-nucleon ($mN$) absorption, described here by 
an added phenomenological potential $V^{(2)}_{K^-}$, and single-nucleon (1$N$) 
absorption encoded in the input $K^-N$ amplitudes. The present approach 
introduces coupling between the corresponding potentials $V^{(2)}_{K^-}$ and 
$V^{(1)}_{K^-}$, and thus it may offer an interesting point of view on this 
topic.

Fig.~\ref{fig:WRW1nonadd} shows the combined IHW-based potential for Ni 
together with its two components. The 1$N$ component is as obtained when there 
is no $mN$ component present, but in the final (combined) potential the 1$N$ 
component is affected by the $mN$ term due to the implicit coupling between 
the two in the SC procedure within the present approach. The non-additivity of 
the two terms is seen, particularly for the imaginary potential. Of greater 
interest is the radial extent of the two terms. At 4.13~fm in this example 
the two terms are equal for the imaginary part of the potential (3.95 fm for 
the real part), with the 1$N$ part clearly dominating at larger radii whereas 
the $mN$ part dominates at smaller radii. 

\begin{table}
\caption{Widths of the `upper' $5g$ state and `lower' $4f$ state in kaonic Ni, 
calculated in the best-fit IHW-based model, see text.}
\label{tab:widths}
\begin{center}
\begin{tabular}{cccccc}
\hline
Re(1$N$) & Im(1$N$) & Re($mN$) & Im($mN$) & $\Gamma_{\rm upper}$ (eV) &
$\Gamma_{\rm lower}$ (keV) \\
\hline
$+$ & $+$ & $+$ & $+$ & 2.7 & 0.97 \\
$+$ & $-$ & $+$ & $+$ & 1.1 & 0.72 \\
$+$ & $+$ & $+$ & $-$ & 2.0 & 1.54 \\
\hline
\end{tabular}
\end{center}
\end{table}

Widths of lower and upper levels in the same kaonic atom could display 
different sensitivities to the 1$N$ and $mN$ absorption terms due to 
differences between the overlaps of the radial atomic wavefunctions 
and the imaginary potential. 
Table \ref{tab:widths} shows, as an example, calculated widths for the upper 
$5g$ and for the lower $4f$ levels in kaonic Ni, calculated by using several 
variants of 1$N$ and $mN$ potential terms of the present model. 
The first row is for the IHW-based potential as obtained from global 
fits to 65 data points across the periodic table. Some idea on the relative 
importance of the 1$N$ and $mN$ absorption terms may be gained by removing 
these in turn. However, one must recall that strong-interaction effects in 
kaonic atoms are extremely non-perturbative. In particular, widths of 
lower states are highly saturated, meaning that the addition of absorption 
causes a reduction of widths, due to the repulsive effect on the 
radial wavefunction \cite{FG99a,FG99b}. 
Starting with the upper level, expecting that the local distortions are 
not too strong, then the second row shows that removing the 1$N$ 
absorption depletes the width by 60\% and the third row shows that 
removing the $mN$ absorption depletes it by only 27\%. 
Additivity of the two terms is approximately respected for the width 
of the upper level. The reverse holds true for the lower level. Removing the 
longer-range 1$N$ absorption reduces the width by only 26\% but removing 
the shorter-range $mN$ absorption almost 
doubles the calculated width in this strongly saturated situation. 

\begin{figure}[thb]
\begin{center}
\includegraphics[width=0.6\textwidth]{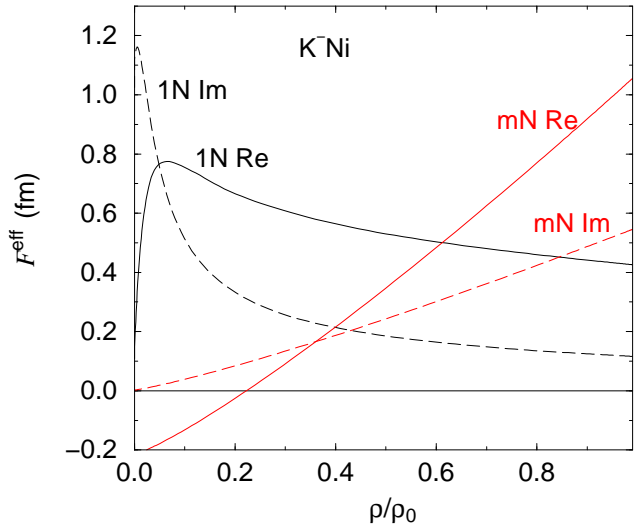}
\caption{Effective amplitudes of the $mN$ term for Ni, 
compared with the 1$N$ amplitudes, see text.}
\label{fig:effampl2N}
\end{center}
\end{figure}

Fig.~\ref{fig:effampl2N} shows the effective amplitude of the $mN$ term 
for Ni compared with the 1$N$ amplitude. This $mN$ amplitude could serve as 
a guide to what a more fundamental theory of multi-nucleon $K^-$ absorption 
and dispersion processes should yield. We note that, based on the IHW 
amplitude, it is possible to obtain a $mN$ phenomenological term whose 
imaginary part is absorptive throughout the nuclear volume. This was not 
the case for the NLO30 model. 

The absorptive $mN$ amplitude shown in Fig.~\ref{fig:effampl2N} is rather 
strong, exceeding the absorptive IHW-based in-medium 1$N$ amplitude for 
densities higher than $\approx$0.4$\rho_0$. A possible comparison with 
experiment is provided by evidence from old emulsion data \cite{VVW77} 
for roughly 1:3 ratio of $mN$:1$N$ absorptivities. This ratio is seen 
in Fig.~\ref{fig:effampl2N} to occur at a density of $\approx$0.2$\rho_0$ 
which is representative of the density range in kaonic atoms where one expects 
absorption to occur. In contrast, the recent chiral-model calculations by 
Sekihara et al. \cite{SSJ12} reach a 1:3 ratio for non-mesonic vs. mesonic 
absorptions at a considerably higher density close to $\rho_0$, implying 
thus that a substantial contribution to multi-nucleon absorption is missing 
in their model. 

\section{Summary} 
\label{sec:sum} 

We have presented fits to kaonic atom data across the whole periodic 
table based on the IHW free-space NLO chiral $K^-N$ amplitudes below 
threshold \cite{IHW12}. The WRW MS procedure \cite{WRW97} was used to 
form in-medium $K^-N$ amplitudes in terms of which a 1$N$ potential 
$V^{(1)}_{K^-}$ is constructed. The strong energy dependence of the 
free-space subthreshold $K^-N$ amplitudes induces substantial density 
dependence in $V^{(1)}_{K^-}$ within the SC calculation of the energy 
parameter $\sqrt s$. This dependence is enhanced further by the 
implicit coupling to a phenomenological $V^{(2)}_{K^-}$ term and 
good fits to the data were reached in this way. It was found, 
in full agreement with part I of this work \cite{FG12} which 
was based on in-medium NLO30 amplitudes due to CS \cite{CS12}, 
that a sizable empirical $mN$ potential was required, both for the 
imaginary part as well as for the real part. By considering in some 
detail the contribution of Im~$V^{(2)}_{K^-}$ to the width of `upper' 
and `lower' states in Ni, we have demonstrated how its relative 
importance develops as one enters the denser regions of the nuclear 
surface and further inward. With a theoretically-based 1$N$ term 
coupled to a phenomenological $mN$ term within a self-consistent 
subthreshold approach, the latter could  guide more theoretical 
work to derive the origin of such a strong multi-nucleon $V^{(2)}_{K^-}$ 
component. Finally, new precision measurements of strong-interaction 
observables for more than a single level on a given target could greatly 
enhance our understanding of the various nuclear absorption processes of 
stopped $K^-$ mesons.

\section*{Acknowledgements} 
Special thanks are due to Ale\v{s} Ciepl\'{y} for providing us with 
a separable-model in-medium extension of the free-space IHW $K^-N$ NLO 
amplitudes, and to Wolfram Weise for stimulating discussions throughout 
the final stages of this work. Financial support by the EU initiative FP7, 
HadronPhysics3 (SPHERE and LEANNIS networks), is gratefully acknowledged.

\section*{Appendix A.~~On coupling $V^{(2)}_{K^-}$ explicitly to 
$V^{(1)}_{K^-}$} 
\label{sec:appendix} 
\renewcommand{\theequation}{A.\arabic{equation}}
\setcounter{equation}{0}
\renewcommand{\thefigure}{A.\arabic{figure}} 
\setcounter{figure}{0} 

Here we wish to present alternative approaches in which the dispersive and 
absorptive nuclear processes are viewed as forming a {\it background} 
potential $V^{(2)}_{K^-}$ that affects meson propagation in between 
scatterings. In the first approach, the incident meson wave $\psi_{\rm inc}$ 
is no longer a free-space plane wave $\exp({\rm i}{\vec k}\cdot{\vec r})$. 
It now satisfies the wave equation 
\begin{equation} 
(\Delta + k^2)\;\psi_{\rm inc}=2{\omega_K}V^{(2)}_{K^-}(\omega_K;\rho)\;
\psi_{\rm inc} \;, 
\label{eq:psi} 
\end{equation} 
which for constant density $\rho$ admits a solution 
$\psi_{\rm inc}=\exp({\rm i}{\vec k}_{\rho}\cdot{\vec r})$ with 
\begin{equation} 
{k_{\rho}}^2=k^2-2{\omega_K}V^{(2)}_{K^-}(\omega_K;\rho) \;. 
\label{eq:krho} 
\end{equation} 

It is straightforward then to revise the MS exposition Eqs.~(\ref{eq:wrw6})
--(\ref{eq:wrw25}) in Section~\ref{subsec:wrw}, replacing the 
incident free-space meson wave $\exp({\rm i}{\vec k}\cdot{\vec r})$ by the 
in-medium incident wave $\exp({\rm i}{\vec k}_{\rho}\cdot{\vec r})$ where 
$k_{\rho}$ is defined by (\ref{eq:krho}). The outcome is that an 
overall potential $V_{K^-}=V^{(1)}_{K^-}+V^{(2)}_{K^-}$ is identified, with 
Eq.~(\ref{eq:wrw26}) for $V^{(1)}_{K^-}$ re-derived under the replacement 
$\xi_k \to \xi_{k_{\rho}}$: 
\begin{equation}  
2{\omega_K}V^{(1)}_{K^-}(\omega_K;\rho)=-4\pi\;\sum_{I}\frac{2I+1}{4}\;
\frac{{\tilde f}_{I}(\sqrt{s_{\rho}})}{1+\frac{1}{4}\xi_{k_{\rho}}
{\tilde f}_{I}(\sqrt{s_{\rho}})\rho}\:\rho \;, 
\label{eq:V1opt} 
\end{equation} 
where $\xi_{k_{\rho}}$, in analogy with (\ref{eq:wrw25}), is given by  
\begin{equation} 
\xi_{k_{\rho}}=\frac{9\pi}{p_F^2}\left( 4\int_0^{\infty}{\frac{dt}{t}\,
\exp(iq_{\rho}t)\,j_1^2(t)} \right) \;, \;\;\;\;\;  q_{\rho}=k_{\rho}/p_F \;, 
\label{eq:xikrho} 
\end{equation} 
with $k_{\rho}$ defined by (\ref{eq:krho}) through the input $V^{(2)}_{K^-}$. 
The integral in (\ref{eq:xikrho}) can be evaluated analytically, resulting in: 
\begin{equation} 
\xi_{k_{\rho}}=\frac{9\pi}{p_F^2}\left[ 1+\frac{{\rm i}q_{\rho}}{2}\left( 
( 1-\frac{{q_{\rho}}^2}{4})\ln(1-\frac{4}{{q_{\rho}}^2})-1\right) \right] \;. 
\label{eq:xikrhofinal} 
\end{equation} 
$V^{(1)}_{K^-}$, Eq.~(\ref{eq:V1opt}), is now explicitly coupled to 
$V^{(2)}_{K^-}$ which enters via (\ref{eq:krho}) and the r.h.s. expression of 
$\xi_{k_{\rho}}$ in Eq.~(\ref{eq:xikrhofinal}). 

An alternative form of considering explicit coupling between $V^{(2)}_{K^-}$ 
and $V^{(1)}_{K^-}$ is motivated by observing that the single-nucleon 
$\bar K N$ scattering amplitudes ${\tilde f}_I$ which enter the r.h.s. of 
Eqs.~(\ref{eq:wrw26}) and (\ref{eq:V1opt}) for $V^{(1)}_{K^-}$ are evaluated 
below the $\bar K N$ threshold. Their imaginary parts are related there 
exclusively to $K^-N\to \pi Y$ reaction processes which do not require 
imposing the Pauli principle in the final $\pi Y$ states. This suggests 
that the MS $V^{(1)}_{K^-}$ is constructed from the real parts of 
${\tilde f}_I$, while the imaginary parts are attached to the background 
potential $V^{(2)}_{K^-}$. We denote the modified $V^{(j)}_{K^-}$ by 
$U^{(j)}_{K^-}$, so that $V_{K^-}=U^{(1)}_{K^-}+U^{(2)}_{K^-}$ as follows: 
\begin{equation} 
2{\omega_K}U^{(1)}_{K^-}(\omega_K;\rho)=-4\pi\:\sum_{I}\frac{2I+1}{4}\;
\frac{{\rm Re}{\tilde f}_{I}(\sqrt{s_{\rho}})}{1+\frac{1}{4}\xi_{k_{\rho}}
{\rm Re}{\tilde f}_{I}(\sqrt{s_{\rho}})\rho}\:\rho  \;, 
\label{eq:U1} 
\end{equation} 
\begin{equation} 
U^{(2)}_{K^-}(\omega_K;\rho)=V^{(2)}_{K^-}(\omega_K;\rho)-\frac{4\pi}
{2{\omega_K}}{\rm i}\:\sum_{I}\frac{2I+1}{4}\;{\rm Im}{\tilde f}_{I}
(\sqrt{s_{\rho}})\rho\;, 
\label{eq:U2} 
\end{equation} 
with $\xi_{k_{\rho}}$ from Eqs.~(\ref{eq:xikrho}) and (\ref{eq:xikrhofinal}), 
and $k_{\rho}$ now defined by 
\begin{equation} 
{k_{\rho}}^2=k^2-2{\omega_K}U^{(2)}_{K^-}(\omega_K;\rho) \;. 
\label{eq:krhomodif} 
\end{equation} 
Similarly to the discussion in Section~\ref{subsec:V2}, it is optional to 
ignore any explicit coupling between $U^{(1)}_{K^-}$ and $U^{(2)}_{K^-}$ by 
using $\xi_{k=0}=9\pi/p_F^2$ for $\xi_{k_{\rho}}$ on the r.h.s. expression 
(\ref{eq:U1}) for $U^{(1)}_{K^-}$. 

\begin{figure}[htb] 
\begin{center} 
\includegraphics[width=0.6\textwidth]{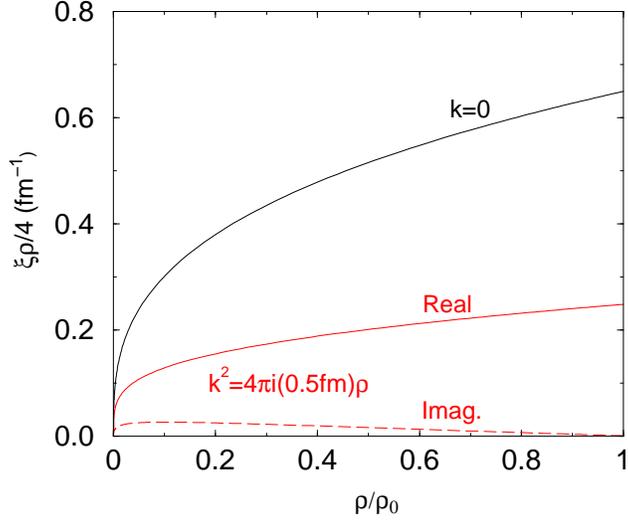} 
\caption{The density dependence of the MS complex expansion parameter 
$\frac{1}{4}\xi_{k_{\rho}}\rho$, with $\xi_{k_{\rho}}$ from 
Eq.~(\ref{eq:xikrhofinal}), for a background absorptive amplitude 
${\tilde{\cal F}}_{\rm bgd}^{\rm eff}={\rm i}\:0.5$~fm. The upper curve shows, 
for comparison, the WRW MS real expansion parameter for no background 
amplitude, i.e. $9\pi\rho/4p_F^2$.} 
\label{fig:ximodels} 
\end{center} 
\end{figure} 

Before closing we demonstrate briefly the additional density dependence 
implied by using $\xi_{k_{\rho}}$, Eq.~(\ref{eq:xikrhofinal}), in the MS 
renormalization on the r.h.s. of Eqs.~(\ref{eq:V1opt}) and (\ref{eq:U1}), 
compared to using $\xi_{k=0}=9\pi/p_F^2$ in the main text. To this end we 
assume a density-independent background (bgd) absorptive amplitude 
${\tilde{\cal F}}_{\rm bgd}^{\rm eff}={\rm i}\:0.5$~fm so that 
${k_{\rho}}^2={\rm i}\:2\pi\rho$ in fm-derived units. 
Fig.~\ref{fig:ximodels} shows the MS complex expansion parameter 
$\frac{1}{4}\xi_{k_{\rho}}\rho$ along with the real parameter 
$9\pi\rho/4p_F^2$ which holds in the limit $k_{\rho}=0$. 
We observe that the imaginary part of the expansion parameter is 
negligible with respect to the real part, also in absolute terms, and that 
the real part is considerably quenched relative to its $k_{\rho}=0$ limit. 
Consequently, the WRW renormalization of the single-nucleon amplitudes 
$\tilde f$ is somewhat weaker than when using $\xi_{k=0}=9\pi/p_F^2$. 
This conclusion holds true also when the balance between real and imaginary 
parts of the expansion parameter $\frac{1}{4}\xi_{k_{\rho}}\rho$ varies, 
depending on the precise complex structure of the assumed background amplitude 
${\tilde{\cal F}}_{\rm bgd}^{\rm eff}$.


\begin{thebibliography}{99}

\bibitem{BFG97} C.J.~Batty, E.~Friedman, A.~Gal, Phys. Rep. 287 (1997) 385. 

\bibitem{FG07} E.~Friedman, A.~Gal, Phys. Rep. 452 (2007) 89. 

\bibitem{CFG11a} A.~Ciepl\'{y}, E.~Friedman, A.~Gal, D.~Gazda, J.~Mare\v{s}, 
Phys. Lett. B 702 (2011) 402. 

\bibitem{CFG11b} A.~Ciepl\'{y}, E.~Friedman, A.~Gal, D.~Gazda, J.~Mare\v{s}, 
Phys. Rev. C 84 (2011) 045206. 

\bibitem{FG12} E.~Friedman, A.~Gal, Nucl. Phys. A 881 (2012) 150. 

\bibitem{CS10} A.~Ciepl\'{y}, J.~Smejkal, Eur. Phys. J. A 43 (2010) 191. 

\bibitem{CS12} A.~Ciepl\'{y}, J.~Smejkal, Nucl. Phys. A 881 (2012) 115. 

\bibitem{SID11} M.~Bazzi, et al., Phys. Lett. B 704 (2011) 113. 

\bibitem{SID12} M.~Bazzi, et al., Nucl. Phys. A 881 (2012) 88.

\bibitem{IHW11} Y.~Ikeda, T.~Hyodo, W.~Weise, Phys. Lett. B 706 (2011) 63. 

\bibitem{IHW12} Y.~Ikeda, T.~Hyodo, W.~Weise, Nucl. Phys. A 881 (2012) 98. 

\bibitem{WRW97} T.~Waas, M.~Rho, W.~Weise, Nucl. Phys. A 617 (1997) 449. 
This MS framework was used in the kaonic quasibound-state 
calculations of Ref.~\cite{WH08}. 

\bibitem{WH08} W.~Weise, R.~H\"{a}rtle, Nucl. Phys. A 804 (2008) 173. 

\bibitem{EE66} M.~Ericson, T.E.O.~Ericson, Ann. Phys. 36 (1966) 323; 
M.~Krell, T.E.O.~Ericson, Nucl. Phys. B 11 (1969) 521. 

\bibitem{FGB94} E.~Friedman, A.~Gal, C.J.~Batty, Nucl. Phys. A 579 (1994) 518.

\bibitem{BFr07} N.~Barnea, E.~Friedman, Phys. Rev. C 75 (2007) 022202(R).

\bibitem{FG99a} E.~Friedman, A.~Gal, Phys. Lett. B 459 (1999) 43. 

\bibitem{FG99b} E.~Friedman, A.~Gal, Nucl. Phys. A 658 (1999) 345. 

\bibitem{VVW77} C.~Vander Velde-Wilquet, J.~Sacton, J.H.~Wickens, D.N.~Tovee, 
D.H.~Davis, Nuovo Cimento 39 A (1977) 538, and references therein.

\bibitem{SSJ12} T.~Sekihara, J.~Yamagata-Sekihara, D.~Jido, 
Y.~Kanada-En'yo, Phys. Rev. C 86 (2012) 065205. 


















\end{thebibliography}
\end{document}